\documentclass[a4paper]{iopconfser}
\usepackage{amsmath,amssymb,amsfonts,mathtools,graphicx,hyperref}
\newcommand*{\doi}[1]{\href{https://doi.org/\detokenize{#1}}{doi: \detokenize{#1}}}

\newcommand{\ms}{\medskip}

\def\l{\mathop{\ell}\nolimits}

\begin{document}

\title{Normalized eigenfunctions of parametrically factored Schr\"odinger equations}

\author{J. de la Cruz and H.C. Rosu 
}

\affil{ 
Instituto Potosino de Investigaci\'on Cient\'{\i}fica y Tecnol\'ogica, Camino a la Presa San Jos\'e 2055,
Col. Lomas 4a Secci\'on, San Luis Potos\'{\i}, 78216 S.L.P., Mexico}\vspace{3mm}

\email{josue.delacruz@ipicyt.edu.mx; ORCID: 0000-0001-5943-5752}
\email{hcr@ipicyt.edu.mx; ORCID: 0000-0001-5909-1945}

\begin{abstract}
The factorizations using the general Riccati solution constructed from a given particular solution by means of the Bernoulli ansatz initiated in 1984 by Mielnik and Fern\'andez C. for the cases of the quantum harmonic oscillator and the radial Hydrogen equation, respectively, are briefly reviewed. The issue of the eigenfunction normalization of the obtained one-parameter Darboux-deformed Hamiltonians is addressed here. 
\end{abstract}

\section{Introduction}

\ms

According to Infeld and Hull \cite{IH51}, the factorization method for Sturm-Liouville differential equations owes its existence primarily to Schr\"odinger, who at the beginning of 1940's discussed the method \cite{Schr40} and also factored in four different ways the hypergeometric differential equation \cite{Schr41}. However, they also mention Weyl \cite{W31} and Dirac \cite{Dir30} as precursors, who, a decade earlier, considered the spherical harmonics with spin and the creation and annihilation operators, respectively, in a similar framework. Dirac himself gave credit to Fock for the latter factoring operators. After the comprehensive paper of Infeld and Hull from 1951, there were only very few works, {\it e.g.}, Crum \cite{C55}, related to the factorization methods that have been published along the next three decades. However, during the 1980's, with the advent of supersymmetric quantum mechanics, the factorizations in their supersymmetric version with the usage of particular Riccati solutions have become the preferred tool for generating isospectral potentials with different degrees of isospectrality. On the other hand, although the factorizations based on general Riccati solutions for the fundamental cases of quantum harmonic oscillator and the Hydrogen atom have been also addressed in the same period in the pioneering papers of Mielnik \cite{M84} and Fern\'andez C. \cite{DF84}, respectively, they really did not enter the mainstream research of supersymmetric quantum mechanics, although one can see chapter 6 in \cite{CKS2001} and references therein and in the review paper \cite{MR2004}. 
In fact, more attention attracted their connection with the version of the Gel'fand-Levitan inverse method developed by Abraham and Moses \cite{AM80}, as mentioned in \cite{M84,DF84} and also noticed by Nieto \cite{N84}. In a parallel line of research, after Matveev \cite{M79a,M79b} revived the Darboux transformations \cite{D1882}, Andrianov {\it et al.} \cite{ABI84a,ABI84b} showed their equivalence with the supersymmetric factorization method and therefore one may also consider the parametric factorizations as a particular kind of parametric Darboux transformations.   

\ms 

The main goal of this short paper, which is dedicated to Professor D.J. Fern\'andez on the occasion of his forty years of remarkable scientific activity, is to revisit the pioneering papers of 1984, pinpointing some details, such as the issue of normalization constants of the parametrically deformed wavefunctions, which has not been mentioned in the original papers. 

\ms

The rest of the paper is structured as follows. In Section 2, the parametric factorization of a Schr\"odinger like equation is presented in general terms. Section 3 is devoted to the application to the quantum harmonic oscillator following \cite{M84}, with a subsection on the normalization constant of the deformed oscillator eigenfunctions. The application to the radial equation of the Hydrogen atom following \cite{DF84}, with a similar subsection on the normalization constants, is presented in Section 4. The Conclusions section ends up the paper. 

\section{The one-parameter factorization}

\ms

Let us consider 
the one-dimensional Schr\"odinger equation 
\begin{equation}\label{Schr}
\left(-D^2+ f(x)-E \right) F=0~,\qquad D^2=\frac{d^2}{dx^2}~,  
\end{equation}
where the Schr\"odinger potential $f(x)$, is an arbitrary $C^2$ function, and also primes will be used in the following for the derivatives of some functions.
We set $E=0$ for the ground state energy, thus turning the equation into a homogeneous differential equation for the ground state wavefunction $F_0$ 
\begin{equation}\label{Schr0}
\left(-D^2+ f(x)\right) F_0=0~.   
\end{equation}
Equation (\ref{Schr0}) can be factored in the form
\begin{equation}\label{Schr1}
(-D+\Phi)(D+\Phi)F_0\equiv \big[-D^2+(\Phi^2-\Phi')\big]F_0=0~, \qquad
\end{equation}
where $\Phi=-F_0'/F_0$ is the negative logarithmic derivative of a solution $F_0$ of (\ref{Schr0}), a.k.a. the seed solution of the Darboux transformation (DT). Comparing (\ref{Schr0}) and (\ref{Schr1}), one obtains the Riccati equation,
\begin{equation}\label{Ric1}
\Phi^2-\Phi'=f(x)~.
\end{equation}

If one knows a Riccati solution $\Phi$ of (\ref{Ric1}), the solution of (\ref{Schr0}) can be obtained from $F_0={\rm exp}(-\int^x \Phi)$.

\medskip

The non-parametric Darboux-transformed equation of (\ref{Schr1}), also known as the supersymmetric partner equation of (\ref{Schr1}), is obtained by reverting the factoring
\begin{equation}\label{SchrFibDT}
(D+\Phi)(-D+\Phi)\hat{F}_0\equiv \big[-D^2+(\Phi^2+\Phi')\big]\hat{F}_0=\big[-D^2+(f+2\Phi')\big]\hat{F}_0=0~.
\end{equation}
The generic interesting fact of the reverted factorizations is that they are not unique, but parametric. 
This is because in the factorization brackets one can use the general Riccati solution in the form of the Bernoulli ansatz $\Phi_g=\Phi+1/u$, where the function $u$ satisfies the first-order differential equation
\begin{equation}\label{SchrFibDTl}
-u'+2\Phi u+1=0
\end{equation}
and not just a particular solution $\Phi$.

Then, it is easy to show that
\begin{equation}\label{SchrFibDTp}
\left(D+\Phi+\frac{1}{u}\right)\left(-D+\Phi+\frac{1}{u}\right)\bar{F}_0
= \big[-D^2+(\Phi^2+\Phi')\big]\bar{F}_0=0~.
\end{equation}
Furthermore, the left hand side of the latter equation can be written as
\begin{equation}\label{SchrFibDTp2}
\big[-D^2+\left(\Phi^2_{g}+\Phi'_{g}\right)\big]\bar{F}_0=0~.
\end{equation}
Therefore, $\bar{F}_0=\hat{F}_0$, and it does not matter if one uses the particular or general Riccati solution.

\ms

The relevant result is obtained only when one reverts back the factorization brackets in the intent to return to the initial equation. Then, one obtains
\begin{equation}\label{SchrFibDTp3}
\big[-D^2+\left(\Phi^2_{g}-\Phi'_{g}\right)\big]\widetilde{{\cal F}}_0=0~.
\end{equation}
Substituting the Bernoulli form of $\Phi_{g}$ in (\ref{SchrFibDTp3}) leads to
\begin{equation}\label{SchrFibDTp4}
\left(-D^2+ f(x)+4\Phi u^{-1}+2u^{-2}\right)\widetilde{{\cal F}}_0=0~.
\end{equation}

The latter equation is actually a one-parameter family of equations having the same Darboux-transformed partner, the running parameter of the family being the integration constant that occurs in Bernoulli's function $1/u$ obtained by the integration of (\ref{SchrFibDTl})
\begin{equation}\label{SchrFibDTp5}
\frac{1}{u}=\frac{e^{-2\int^x \Phi(x')dx'}}{\gamma +\int^x e^{-2\int^{\bar{x}} \Phi(x')dx'}d\bar{x}}=\frac{d}{dx}\ln\left(\gamma +\int^x e^{-2\int^x \Phi(x')dx'}dx\right)~.
\end{equation}
It is easy to show that equation (\ref{SchrFibDTp4}) differs from the initial equation (\ref{Schr}) by
\begin{equation}\label{Darbdef}
4\Phi/u+2/u^2=2\frac{1}{u}\frac{d}{dx}\ln \left(\frac{1}{u}\right)~,
\end{equation}
which is the additive Darboux deformation of $f(x)$ introduced by the parametric DT.
Moreover, the solutions $\widetilde{{\cal F}}_0$ of the parametric Darboux-transformed equations are related to the undeformed solutions $F_0$ as follows
\begin{equation}\label{SchrFibDTp6}
\widetilde{{\cal F}}_0=e^{-\int^x \Phi_{g} ds}=e^{-\int^x \Phi ds}e^{-\int^x\frac{d}{dx}\ln\left(\gamma +\int^x e^{-2\int^x \Phi(s)ds}\right)}
=\frac{F_0}{\gamma +\int^x F_0^2(s)ds}~.
\end{equation}

Before ending this section, we mention that including a scalar scaling $c$ in front of $D^2$ can be manipulated by democratically distributing $\sqrt{c}$ in front of each of the factoring brackets $(\pm D+\Phi)$. 
The only change is the scaling by $c$ of the eigenspectrum of the given Schr\"odinger problem. Therefore, in the following we will use the usual quantum-mechanical scaling $1/\sqrt{2}$ in front of the factoring operators.  

\section{The Quantum Harmonic Oscillator ($\Phi=x$)}

\ms

For this case, the factorization brackets are similar to Fock's creation and annihilation operators
$a=\frac{1}{\sqrt{2}}\left( \frac{d~}{dx}+x \right)$ and its adjoint conjugate $a^\dagger$, which factorize the Hamiltonian $H_{qho}=-\frac{1}{2}\frac{d^2~}{dx^2}+\frac{1}{2} x^2$ as
\begin{equation}\label{m1}
aa^\dagger=H+\frac{1}{2}~, \quad {\rm and} \quad a^\dagger a=H-\frac{1}{2}~,
\end{equation}
are replaced by the new operators $b=\frac{1}{\sqrt{2}}\left( \frac{d~}{dx}+\beta(x) \right)$, and its adjoint conjugate $b^\dagger$, requiring that
\begin{equation}\label{m2}
bb^\dagger=H+\frac{1}{2}~, \quad b^\dagger b = \widetilde{H}-\frac{1}{2}~.
\end{equation}
In order to satisfy this, $\beta(x)$ is found to satisfy the Riccati equation
\begin{equation}\label{m3}
\beta'+\beta^2=1+x^2
\end{equation}
and the reversed factorization $b^\dagger b$ provides the one-parameter isospectral Hamiltonian with the one-parameter contribution to the potential
\begin{equation}\label{m4}
\widetilde{H}=H+\frac{d~}{dx}\left[ \frac{e^{-x^2}}{\gamma+\int_{-\infty}^x e^{-x'^2}dx'} \right]~,
\end{equation}
where the real parameter $\gamma$ is required to be $|\gamma|>\frac{1}{2}\sqrt{\pi}$ if we want to avoid singular potentials.

From (\ref{m4}) and (\ref{SchrFibDTp6}) with $F_0=e^{-\frac{x^2}{2}}$ and $\widetilde{{\cal F}}_0=\widetilde{\Psi}_0(x;\gamma)$, one obtains
\begin{equation}\label{m4bis}
\widetilde{V}(x;\gamma)=\frac{x^2}{2}+\frac{d}{dx}\left(\frac{e^{-x^2}}{\gamma+\int_{-\infty}^{x}e^{-x'^2}dx'}\right)~,\quad 
\widetilde{\Psi}_0(x;\gamma)=\frac{e^{-x^2/2}}{\gamma+\int_{-\infty}^{x}e^{-x'^2}dx'}~,
\end{equation}
respectively. In Fig.~\ref{fig1pho}, one can observe plots of singular and regular one-parameter deformed counterparts of the harmonic oscillator potential and ground state for $\gamma=\mp 0.5$, respectively.

\begin{center}
\begin{figure}[h!] 
\includegraphics[width=0.50\linewidth] {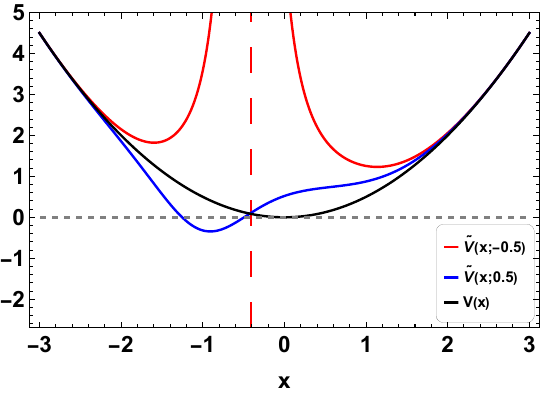}   
\includegraphics[width=0.50\linewidth]{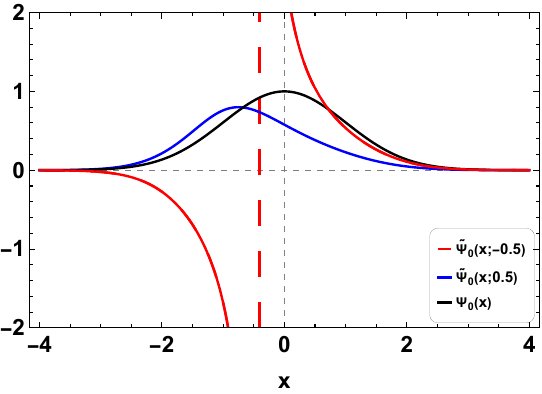}
\caption{Singular (red) and regular (blue) Mielnik potentials and ground state wavefunctions $\widetilde{\Psi}_0(x;\gamma)$
for $\gamma=\mp 0.5$. The undeformed harmonic oscillator potential and its unnormalized ground state Gaussian wavefunction are also included in black color for comparison as it is done with the appropriate undeformed cases in all the plots of the paper.}
\label{fig1pho}
\end{figure}
\end{center}

\subsection{The normalization constant of the parametrically-deformed oscillator wavefunctions}

\medskip

To determine the normalization constant of the one-parameter Darboux-deformed oscillator ground state, we write 
the normalization integral
\begin{equation}\label{mn2}
\int_{-\infty}^\infty N_0^2\frac{e^{-x'^2}dx'}{(\gamma+\int_{-\infty}^x e^{-x'^2}dx')^2}=1~,
\end{equation}
which, by introducing the change of variable $X(x)=\int_{-\infty}^x e^{-x'^2}dx'$,  becomes
\begin{equation}\label{mn3}
\int_0^{\sqrt{\pi}}N_0^2\frac{dX}{(\gamma+X)^2}=1~.
\end{equation}
Effecting the integral gives
\begin{equation}\label{mn4}
N_0\equiv N_0(\gamma)=\sqrt{\gamma\left(\frac{\gamma}{\sqrt{\pi}}+1\right)}~.
\end{equation}
This shows that $\gamma$ should not be in $[-\sqrt{\pi},0]$.
In Fig.~2, we present plots of two regular Darboux-parametric potentials and the corresponding normalized eigenfunctions.
\begin{center}
\begin{figure}[h!] 
\includegraphics[width=0.50\linewidth]{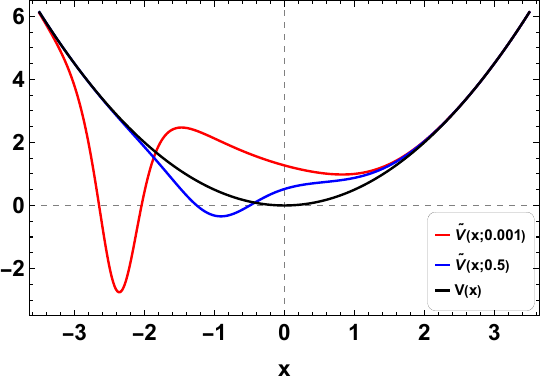}   
\includegraphics[width=0.50\linewidth]{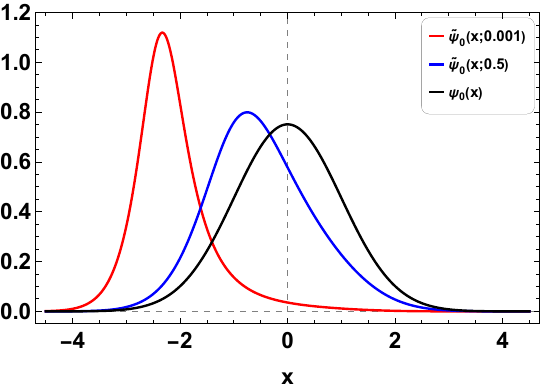}    
\caption{Regular Mielnik potentials and corresponding normalized ground state wavefunctions $\tilde{\psi}_0(x;\gamma)=
N_0(\gamma)\widetilde{\Psi}_0(x;\gamma)$ for $\gamma=0.5$ and $1$,
red and blue, respectively.}
\label{fig1pho}
\end{figure}
\end{center}
Another interesting fact about the normalization of the Darboux-deformed ground states is that there is a pair of parameters 
for which the normalization constant is exactly that of the original ground state, despite the deformed wavefunctions not being Gaussians. In the case of the harmonic oscillator, this pair is given by
 \begin{equation}\label{pair}
 \gamma_{\pm}= \frac{1}{2}\left(-\sqrt{\pi} \pm \sqrt{\pi+4}\right)
 \end{equation}
 for which $N_0(\gamma_{\pm})=\pi^{-1/4}$. These cases are presented in Fig.~\ref{figpmg}, where one can notice the symmetries
 $\tilde{V}(x;\gamma_-)=\tilde{V}(-x;\gamma_+)$ and $\tilde{\psi}(x;\gamma_-)=-\tilde{\psi}(-x;\gamma_+)$.
 For other values of the parameter, the normalization constant can be very big. 
 For example, for $ \gamma_{m}=m\sqrt{\pi}$, one obtains $N_0(m)=\sqrt{m(m+1)}\pi^{1/4}$.
\begin{center}
\begin{figure}[h!] 
\includegraphics[width=0.50\linewidth] {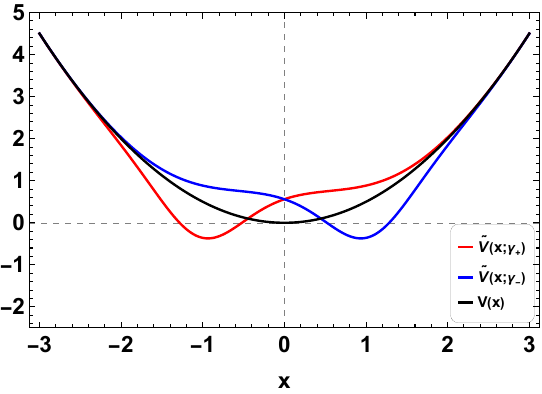}  
\includegraphics[width=0.50\linewidth] {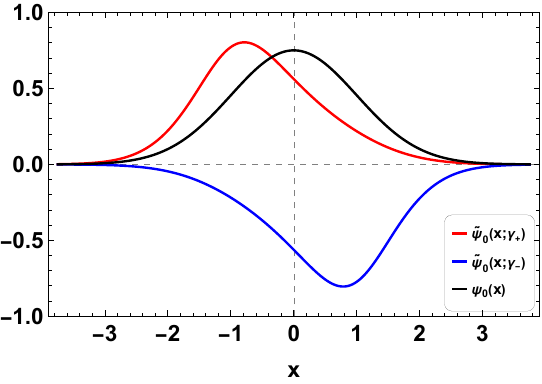} 
\caption{The one-parameter deformed oscillator potentials and corresponding normalized ground state wavefunctions for $\gamma_{\pm}$ from (\ref{pair}).}
\label{figpmg}
\end{figure}
\end{center}

\ms

\section{The Radial Hydrogen Equation ($\Phi=\frac{\ell}{r}-\frac{1}{\ell}$)}

\ms

The radial Schr\"odinger equation for the Hydrogen atom may be written as the following  Sturm-Liouville eigenvalue problem

\begin{equation}\label{hamhyd}
H_{\ell} \, R(r)=\frac{1}{r}\left[ -\frac{d^2 ~}{dr^2}+\frac{\ell ( \ell + 1)}{r^2}-\frac{2}{r}\right] r\,R(r) = \lambda_n R(r)~,
\end{equation}
where $H_{\ell}$ is the radial Hamiltonian operator and the used units are $m=\hbar=e=4 \pi\varepsilon_{0} = 1$, together with $Z$=1.

\ms

With this convention, the energy eigenvalues and the eigenfunctions can be written in the form
\begin{equation}\label{rnls}
\lambda_n = -\frac{1}{n^2} \, , \ \ \
R_{n,\ell}(r)=C_{n,\ell} 
\ \left(\frac{2r}{n}\right)^\ell \, e^{-\frac{r}{n}}L^{2\ell+1}_{n-\ell-1}\left(\frac{2r}{n}\right)~,
\end{equation}
where $C_{n,\ell}$ are the normalization constants given by
\begin{equation}\label{rnls1}
C_{n,\ell} =\frac{2}{n^2}\sqrt{\frac{(n-\ell-1)!}{(n+\ell)!}}~,
\end{equation}
$n\in \mathbb{N}$, $n>\ell$, and $L^{2\ell+1}_{n-\ell-1}(2r/n)$ are derivatives of order $2\ell+1$ of the Laguerre polynomials of degree $n-\ell-1$, a.k.a. associated Laguerre polynomials. The polynomial degree $n_r=n-\ell-1$
provides the number of nodes of the eigenfunctions in the radial direction. In particular, we are interested in 'ground-state'
wavefunctions for which $n=\ell+1$.

\ms

The ladder operators of Infeld and Hull provide the following factorizations of the radial H operator (\ref{hamhyd}) 
\begin{align}
a_\ell^+ a_\ell^-=& H_\ell+\frac{1}{\ell^2}~,
\label{IHF1}\\
a_{\ell+1}^- a_{\ell+1}^+=& H_\ell+\frac{1}{({\ell+1})^2}\ , \label{IHF2}
\end{align}
where the Infeld-Hull ladder operators act on the radial functions as follows:
\begin{align}
a^-_\ell R_{n,\l}&=
\frac{1}{r} \left( \frac{d}{dr} + \frac{\ell}{r}-\frac{1}{\ell} \right) r \, R_{n,\ell}
= c_{n\ell} \,
R_{n,\ell-1}~,\label{amenos}\\
a^+_{\ell}R_{n,\ell-1}&=
\frac{1}{r} \left( -\frac{d}{dr} + \frac{\ell}{r}-\frac{1}{\ell} \right) r R_{n,\ell-1}
= c_{n\ell} \, R_{n,\l}~,\label{amas}
\end{align}
where $c_{n\ell}=\frac{\sqrt{(n-\ell)(n+\ell)}}{n \l}$.

\bigskip

Using Mielnik's factorization approach, Fern\'andez C. introduced the following modified Infeld-Hull 
raising and lowering operators \cite{DF84}
\begin{equation} \label{facdf}
\begin{split}
&A^{+}_{\ell} = \frac{1}{r} \left( -\frac{d}{dr} + \beta_{\ell} \right) r~, \quad A^{-}_{\ell} = \frac{1}{r} \left( \frac{d}{dr} + \beta_{\ell} \right) r~, \\ 
&A^{+}_{\ell}   A^{-}_{\ell} =  H_{\ell} + \frac{1}{\ell^2}~, \qquad A^{-}_{\ell}A^{+}_{\ell}= \widetilde{H}_{\ell-1}+\frac{1}{{\ell}^2}~,
\end{split}
\end{equation}
where
\begin{equation}\label{betaell}
\beta_\ell = \frac{\ell}{r}-\frac{1}{\ell}+\frac{r^{2\ell}e^{-2r/\ell}}
{\gamma_{\ell}-\int_0^{r}y^{2\ell}e^{-2y/\ell}dy}~,
\end{equation}
obtained as the general solution of the Riccati equation
\begin{equation}\label{ricbetaell}
-\beta'_\ell+\beta_\ell^2=-\frac{2}{r}+\frac{\l(\l+1)}{r^2}+\frac{1}{l^2}~.
\end{equation}

\bigskip

The parametric radial Hydrogen potentials and ground state wavefunctions (for the eigenvalue $-1/\ell^2$) are given by\\

\begin{equation}\label{Vell}
\widetilde V_{\ell-1}(r;\gamma)= -\frac{2}{r}+\frac{\l(\l-1)}{r^2}+\frac{d}{dr}
\left(\frac{2r^{2\ell}e^{-2r/\ell}}{\gamma_{\ell}-\int_0^{r}y^{2\ell}e^{-2y/\ell}dy}\right)~,
\end{equation}
\begin{equation}\label{Rell}
\widetilde{R}_{\ell,\ell-1}=\frac{r^{\ell-1}e^{-r/\ell}}{\gamma_\ell-\int_0^r r'^{2\ell}e^{-2r'/\ell}dr'}~,
\end{equation}
where
\begin{equation}\label{Hsing}
\gamma_{\ell} >(2\ell)!\left(\frac{\ell}{2}\right)^{2\ell+1}~,
\ \ \mbox{or} \ \  \gamma_{\ell} <0~,
\end{equation}
to avoid singularities.

\bigskip

\subsection{Normalization constants for $\widetilde{R}_{\ell,\ell-1}$}

\medskip

In this case, the normalization integral is
\begin{equation}\label{IntNell1}
\int_0^\infty N_l^2 \widetilde{R}_{\ell,\ell-1}^2 r^2dr \equiv 
\int_0^\infty N_l^2 \frac{r^{2\ell}e^{-2r/\ell}}{\left(\gamma_\ell-\int_0^r r'^{2\ell}e^{-2r'/\ell}dr'\right)^2}dr=1~.
\end{equation}
The change of variable similar to the case of harmonic oscillator 
\begin{equation}\label{IntNell2}
\Gamma_\ell(r)=\int_0^r r'^{2\ell}e^{-2r'/\ell}dr'
\end{equation}
requires the calculation of $\Gamma_\ell(\infty)=\Gamma_\ell$, which can be performed through
recursive integrations by parts providing the result:
\begin{equation}\label{Gell}
\Gamma_{\ell} =(2\ell)!\left(\frac{\ell}{2}\right)^{2\ell+1}~.
\end{equation}
The normalization constant is then given by
\begin{equation}\label{Nell}
N_\ell=\sqrt{\gamma_\ell\left(\frac{\gamma_\ell}{\Gamma_{\ell}}-1\right)}~,
\end{equation}
and $\gamma_\ell$ can take any arbitrary negative values, while if taken positive, should be strictly greater than $\Gamma_\ell$, confirming (\ref{Hsing}).

\ms

Thus:

\ms

For $\ell=1$, i.e., for $\widetilde{R}_{1,0}$, we have $N_1=\sqrt{\gamma_1\left(4\gamma_1-1\right)}$, and one should take $\gamma_1>1/4$.

\ms

For $\ell=2$, i.e., for $\widetilde{R}_{2,1}$, we have $N_2=\sqrt{\gamma_2\left(\frac{\gamma_2}{24}-1\right)}$,
and one should take $\gamma_2>24$, and so forth.

\ms

\begin{figure}[h!] 
\begin{center}
\includegraphics[width=0.45\textwidth]{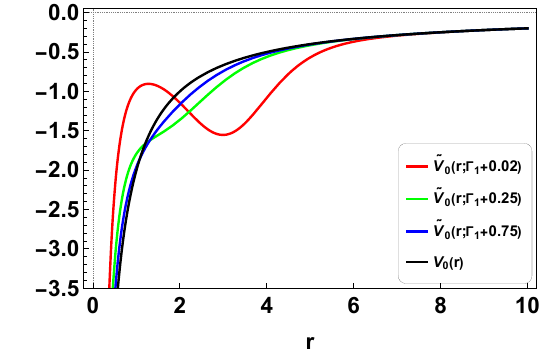} 
\includegraphics[width=0.45\textwidth]{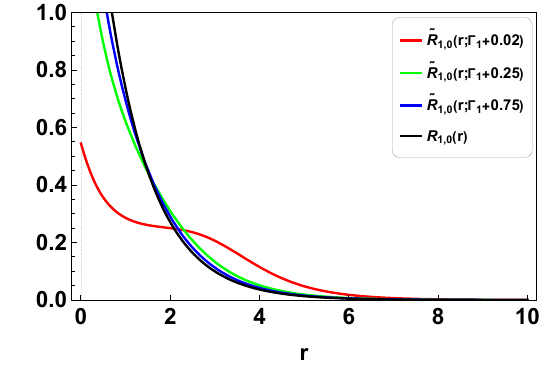}\\           
\includegraphics[width=0.45\textwidth]{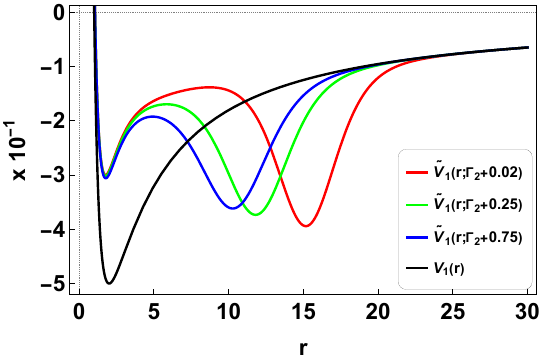}       
\includegraphics[width=0.45\textwidth]{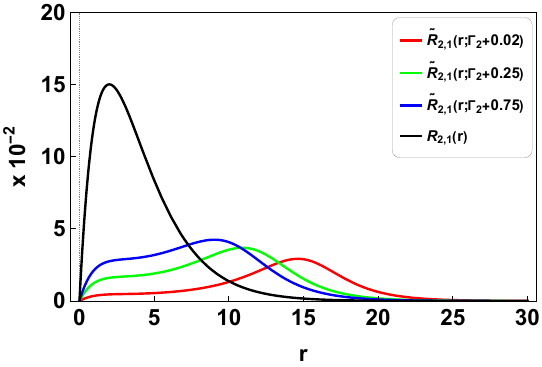}\\           
\includegraphics[width=0.45\textwidth]{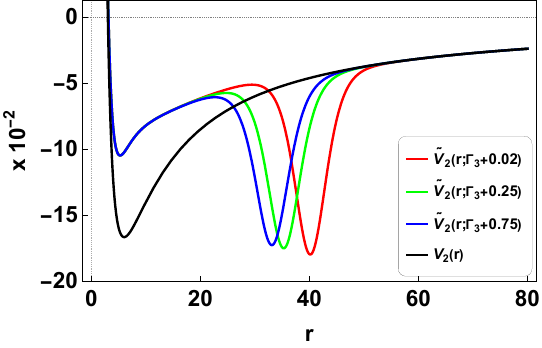}             
\includegraphics[width=0.45\textwidth]{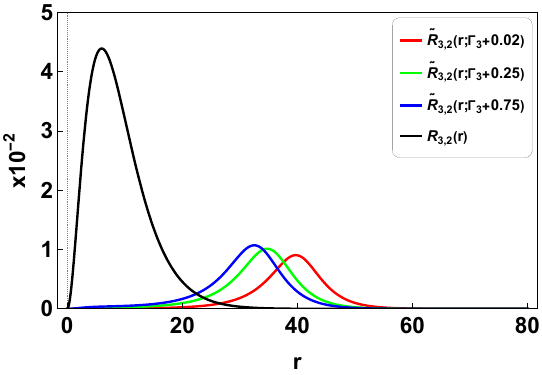}       
\caption{Parametric Hydrogen radial potentials and corresponding normalized one-parameter $\tilde{R}_{\ell,\ell-1}(r;\gamma)$ eigenfunctions for $\ell=1,2,3$.}
\label{figHparam}
\end{center}
\end{figure}

The one-parameter potentials and the corresponding normalized eigenfunctions are presented in Fig.~\ref{figHparam} for $\ell=1,2,3$ together with the undeformed cases for comparison. At high values of $\gamma_\ell$ the supplementary pockets of the deformed potentials tend towards the nondeformed Coulombian well with centrifugal barrier, or without it for the case $\ell=1$, corresponding to the eigenvalue $-1/\ell^2$ as expected.

\ms

\begin{figure}[h!] 
\begin{center}
\includegraphics[width=0.45\textwidth]{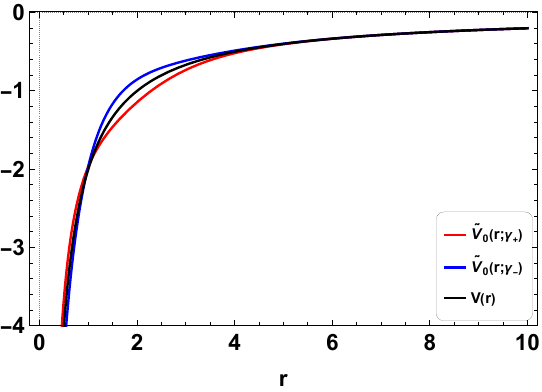}
\includegraphics[width=0.45\textwidth]{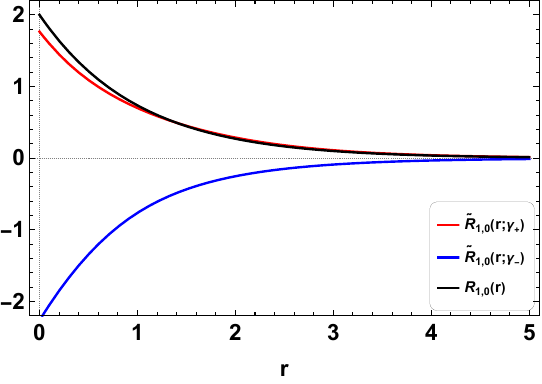}\\
\includegraphics[width=0.45\textwidth] {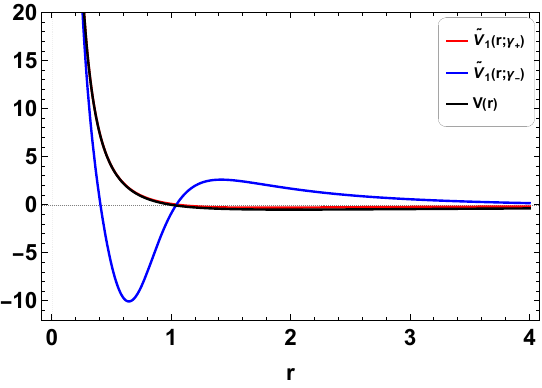} 
\includegraphics[width=0.45\textwidth]{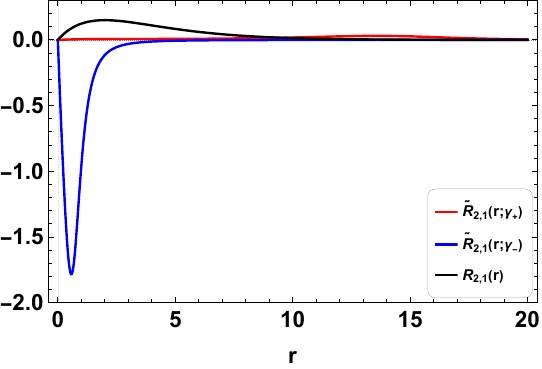} 
\caption{The pairs of  Hydrogen deformed radial potentials $\tilde{V}_{\ell-1}$ and corresponding deformed wavefunctions $\tilde{R}_{\ell,\ell-1}(r;\gamma)$ having the same normalization constant as the undeformed wavefunctions, for $\ell=1,2$.}
\label{figg+-}
\end{center}
\end{figure}

If we ask the deformed wavefunctions to have the same normalization as the undeformed wavefunctions, one can obtain from 
(\ref{rnls1}) that 
$$
C_{\ell,\ell-1}\equiv C_\ell =\frac{2}{l^2\sqrt{(2l-1)!}}~,
$$ 
which then provides a pair of deformation parameters given by the roots of the equation
\begin{equation}\label{Gell}
\gamma^2-\Gamma_\ell \gamma-\Gamma_\ell C_\ell^2=0~.
\end{equation}
For example, from the latter equation, one obtains 
\begin{equation}\label{Gell1}
\gamma_\pm=\frac{1}{8}\left(1\pm \sqrt{65}\right) \qquad {\rm and} \qquad \gamma_\pm=2\sqrt{6}\left(\sqrt{6}\pm \sqrt{6+\frac{1}{24}}\right)
\end{equation}
for $\ell=1$ and $\ell=2$, respectively. Plots of the deformed potentials and normalized wavefunctions for these two cases are presented in Fig.~\ref{figg+-}.

\bigskip
\newpage

\section{Conclusions}

We have revisited the pioneering papers of Mielnik and Fern\'andez that opened the research of parametric families of isospectral potentials in supersymmetric quantum mechanics. The uncovered issue of normalization of the parametric deformed ground state wavefunctions is discussed in some detail. It is found that there exist pairs of deformation parameters (one positive, the other negative) for which the deformed wavefunctions have the same normalization constant as the undeformed wavefunctions.  

\bigskip

{\bf Acknowledgment}

The second author wishes to thank the organizers of the workshop honoring Professor David J. Fern\'andez C. for the enjoyable days of the event. Thanks also to Professors David Fern\'andez and Asim Gangopadhyaya for some clarifying remarks.

\end{document}